\documentclass[noshowpacs,amsmath,twocolumn,superscriptaddress,10pt,aps,prb]{revtex4-1} 
\usepackage{setspace}
\usepackage{amsmath}
\usepackage{breqn}
\usepackage{graphicx}
\usepackage[nearskip,margin = 0pt,caption=false]{subfig}

\usepackage{verbatim}
\usepackage{amsfonts}
\usepackage{amssymb}
\usepackage{epstopdf}
\usepackage{lipsum}
\usepackage{siunitx}
\usepackage{xcolor}
\usepackage{array}
\usepackage{gensymb}
\usepackage[normalem]{ulem}
\usepackage[resetlabels,labeled]{multibib}
\DeclareGraphicsExtensions{.pdf,.eps,.png,.jpg,.mps}
\usepackage{etoolbox}

\usepackage{mathtools}

\usepackage{lineno}
\begin{document}
\title{Octave-spanning Kerr soliton frequency combs in dispersion- and dissipation-engineered lithium niobate microresonators}
\author{Yunxiang Song$^{1,*}$, Yaowen Hu$^1$, Xinrui Zhu$^1$, Kiyoul Yang$^{1,*}$, Marko Lon\v{c}ar$^{1,*}$\\
\vspace{+0.05 in}
$^1$John A. Paulson School of Engineering and Applied Sciences, Harvard University, Cambridge, MA, USA.\\
$^*$Corresponding authors: ysong1@g.harvard.edu, kiyoul@seas.harvard.edu, loncar@g.harvard.edu\\ 
}

\begin{abstract}
\noindent 
\color{black} 

\textbf{Dissipative Kerr solitons from optical microresonators, commonly referred to as soliton microcombs, have been developed for a broad range of applications, including precision measurement, optical frequency synthesis, and ultra-stable microwave and millimeter wave generation, all on a chip. An important goal for microcombs is self-referencing, which requires octave-spanning bandwidths to detect and stabilize the comb carrier envelope offset frequency. Further, detection and locking of the comb spacings are often achieved using frequency division by electro-optic modulation. The thin-film lithium niobate photonic platform, with its low loss, strong second- and third-order nonlinearities, as well as large Pockels effect, is ideally suited for these tasks. However, octave-spanning soliton microcombs are challenging to demonstrate on this platform, largely complicated by strong Raman effects hindering reliable fabrication of soliton devices. Here, we demonstrate entirely connected and octave-spanning soliton microcombs on thin-film lithium niobate. With appropriate control over microresonator free spectral range and dissipation spectrum, we show that soliton-inhibiting Raman effects are suppressed, and soliton devices are fabricated with near-unity yield. Our work offers an unambiguous method for soliton generation on strongly Raman-active materials. Further, it anticipates monolithically integrated, self-referenced frequency standards in conjunction with established technologies, such as periodically poled waveguides and electro-optic modulators, on thin-film lithium niobate.}
\end{abstract}

\maketitle
 
\begin{figure*}[t!]
\centering
\includegraphics[width=\linewidth]{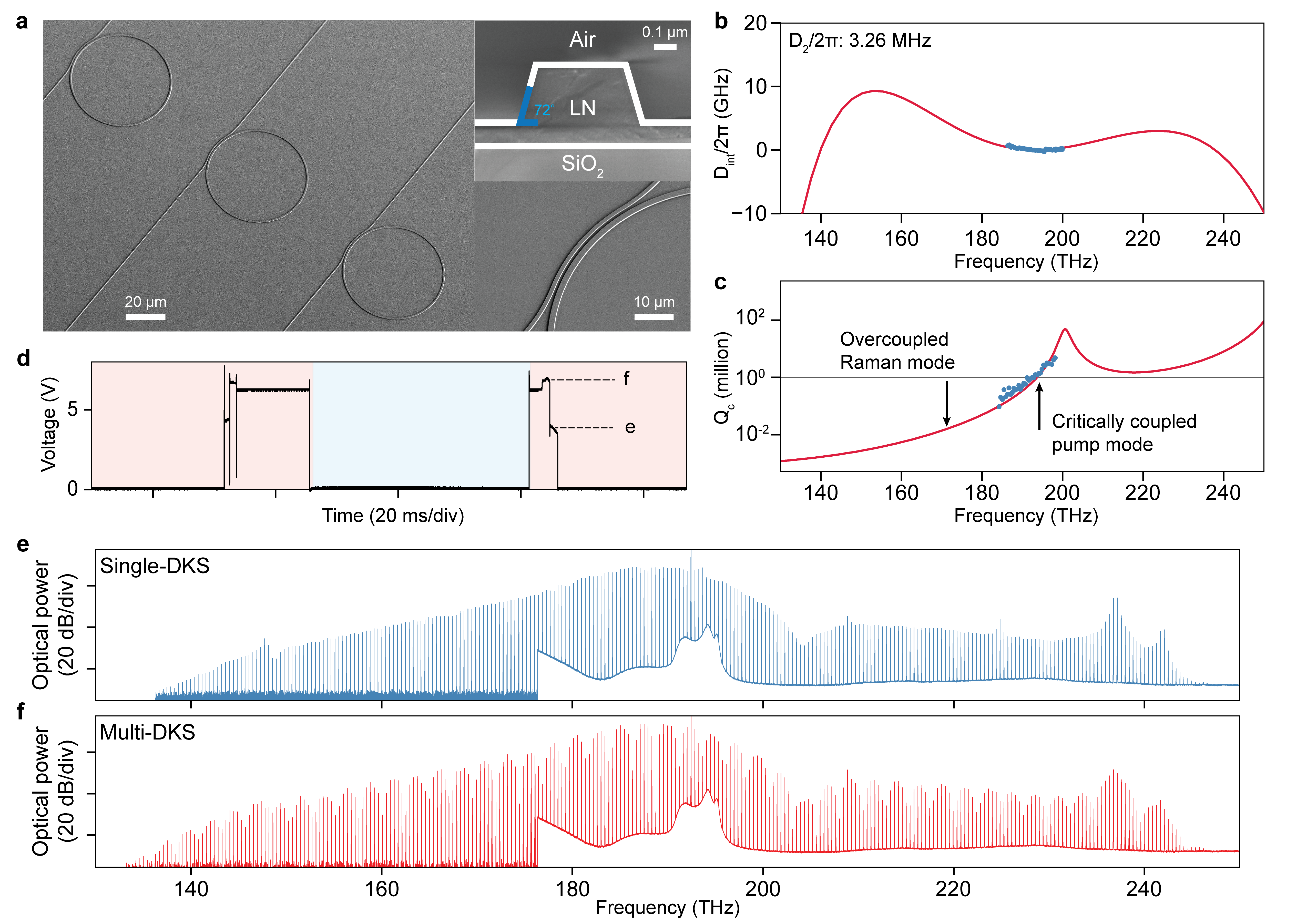}
\renewcommand\figurename{Fig.}
\caption{\label{fig:Fig1}\textbf{DKS device and operation.} \textbf{a}, Scanning electron microscope images of microring array (left), microring waveguide cross section (right, top), and zoom-in of a pulley coupler used for dissipation engineering (right, bottom). The white outlines around the waveguide cross section are a guide to the eye. A sidewall angle of $72\degree$ is achieved through an iterative dry/wet etching method, which provides an additional control knob to engineer the microring dispersion. \textbf{b}, Simulated (solid red line) and measured (blue dots) integrated dispersion ($D_{\text{int}}$) of a 50 $\mu$m-radius microring, assuming bus waveguide width 0.85 $\mu$m, microring waveguide width 1.62 $\mu$m, coupling gap 1.06 $\mu$m, pulley interaction angle $45\degree$, and waveguide height 0.33 $\mu$m. For definition of these parameters and detailed comparison between designed and fabricated devices, see Supplementary information. \textbf{c}, Simulated coupling quality factor $Q_c$ for the same parameters used in \textbf{b}. Measured data (blue dots) is obtained by extracting  $Q_c$ from resonances in the microring transmission spectrum. Black line labels $Q_i\sim$ 1 million. \textbf{d}, Comb power as a photodetector voltage when the pump is swept across a microring resonance, from red to blue side of resonance and back (background color). Two discrete steps marked by black dashed lines correspond to single- and multi-DKS states, indicating bidirectional access under a pump frequency sweep speed of 10 Hz. \textbf{e}, Single-DKS and \textbf{f}, multi-DKS spectra corresponding to steps in \textbf{d}. The comb spacing is about 410.15 GHz and the comb span is about 110 (112.4) THz for the single-DKS (multi-DKS) state. The dips in both DKS spectra around 203 THz is in good agreement with the $Q_c$ peak around $201.5$ THz, as predicted in Fig. 1c.}
\end{figure*}




\section{Introduction}
\color{black} 
Integrated optical frequency combs\cite{pasquazi2018micro,gaeta2019photonic,fortier201920,diddams2020optical,chang2022integrated}, based on microresonator dissipative Kerr solitons (DKS)\cite{herr2014temporal,kippenberg2018dissipative}, have shown promise in providing portable and efficient solutions to a variety of applications, including stable optical\cite{spencer2018optical}, millimeter wave\cite{tetsumoto2021optically}, and microwave\cite{li2014electro,liu2020photonic,yao2022soliton,zhao2024all,kudelin2024photonic,sun2024integrated} frequency generation, precision spectroscopy\cite{suh2016microresonator,picque2019frequency}, astrophysical spectrometer calibration\cite{obrzud2019microphotonic,suh2019searching}, and massively parallel communications\cite{marin2017microresonator,jorgensen2022petabit,yang2022multi,shu2022microcomb,rizzo2023massively}, computing\cite{feldmann2021parallel,xu202111,bai2023microcomb}, and ranging\cite{riemensberger2020massively}. These technological developments motivate the exploration of fully stabilized microcombs, which can provide low-noise and phase coherent frequencies over broad bandwidths, on a chip. Achieving such microcombs critically relies on the detection and stabilization of both the comb carrier envelope offset frequency and the comb spacing\cite{udem2002optical}. Octave-spanning bandwidths coupled with second harmonic generation is required for self-referencing of the carrier envelope offset frequency, and electro-optic modulation is used to convert terahertz comb spacings, typical to octave-spanning DKS, down to the gigahertz (GHz) regime. While the Si$_3$N$_4$ and AlN photonic platforms have realized octave-spanning DKSs\cite{li2017stably,pfeiffer2017octave,liu2021aluminum,weng2021directly}, full stabilization of them remains challenging. Current stabilization efforts still require off-chip lasers\cite{brasch2016photonic,spencer2018optical,liu2021aluminum}, frequency doublers\cite{brasch2017self,spencer2018optical,newman2019architecture}, cascaded chains of bulk electro-optic modulators\cite{brasch2016photonic,drake2019terahertz,moille2023kerr}, and dual comb sources\cite{spencer2018optical,newman2019architecture}, due to the small Pockels effect in AlN and the absence of intrinsic second-order nonlinearity in Si$_3$N$_4$.

The ultra-low loss thin-film lithium niobate (TFLN) photonic platform\cite{zhang2017monolithic,zhu2021integrated,boes2023lithium,zhu2024twenty} offers a promising solution for realizing fully stabilized microcombs, featuring strong electro-optic effect\cite{wang2018integrated,hu2021chip,xu2022dual,hu2024integrated} for dividing large comb spacings, efficient $\chi^{(2)}$ interaction\cite{wang2018ultrahigh,jankowski2020ultrabroadband,mckenna2022ultra} for frequency doubling, and large $\chi^{(3)}$ Kerr effect\cite{wang2019monolithic} for broadband DKS generation. However, while high-speed electro-optic modulation and efficient second harmonic generation are established on TFLN, an entirely connected and octave-spanning DKS is hard to realize. The primary challenge lies in the low threshold Raman lasing driven by high Raman gain over significant bandwidths\cite{yu2020raman,zhao2023widely}, which inhibits DKS formation and drastically complicates systematic fabrication and testing of DKS devices\cite{gong2019soliton}. Previously, TFLN has been used to demonstrate uniquely self-starting and bidirectionally accessible DKS\cite{he2019self,gao2023compact}, dispersion-engineered DKS\cite{zhu2023passively} spanning up to four-fifths of an octave\cite{gong2020near}, photorefraction-enabled free-running DKS\cite{wan2023photorefraction}, breather DKS\cite{lu2023two}, gain-empowered DKS\cite{yang20231550}, monolithic DKS and resonant electro-optic frequency combs\cite{gong2022monolithic}, electro-optically tunable microwave-rate DKS\cite{he2023high}, and electro-optically modulated DKS\cite{song2024hybrid}. Stimulated Raman scattering (SRS) has been suppressed by using 2 $\mu$m light to pump the DKS, though only narrowband combs were achieved in this approach due to dispersion engineering challenges. Other SRS suppression methods based on pulley and self-interference couplers have also been explored, with limited success. For example, the former led to an octave-spanning DKS but suffered from severe undercoupling over a large DKS bandwidth\cite{He:21}, while the latter has not resulted in broadband DKS nor single-DKS states\cite{gong2020photonic}. Despite numerous milestones, conclusive design rules for SRS suppression that support entirely connected and octave-spanning DKS spectra, as well as the realization of DKS devices in a high yield fashion, remains outstanding.

Here, we demonstrate octave-spanning soliton microcombs on TFLN and provide detailed guidelines for achieving a high yield of DKS supporting devices, over various design parameters. We show that octave-spanning DKS can be realized by systematic dispersion engineering and effective SRS suppression using two different methods: (i) the free spectral range (FSR) control method that relies on precise management of the microresonator FSR and (ii) the dissipation engineering method that creates strongly frequency-dependent microresonator to bus waveguide coupling. While the FSR control method can result in an entirely connected DKS spanning 131-263 THz with a comb spacing of 658.89 GHz, it places stringent constraints on microresonator FSRs due to the large Raman gain bandwidth $\Gamma\sim558$ GHz\cite{basiev1999raman,ridah1997composition}. Further, we find that in the FSR$\gtrsim\Gamma$ regime, DKS generation is not deterministic and device yield worsens as the FSR decreases. Conversely, the dissipation engineering method is robust against FSR variations and consistently provides high yield: more than 88 \% of resonance modes across 94 devices support DKS states. Using this method, we also generate entirely connected DKS spanning 126-252 THz with a comb spacing of 491.85 GHz.

\section{Results}
\noindent\textbf{DKS on TFLN}\\
In our work, a DKS is initiated by a red-detuned, continuous wave (CW) pump coupled to a microresonator, where self-phase modulation of the pump is counter-acted by anomalous dispersion, and the microresonator loss is compensated by third-order parametric gain. These double-balancing conditions give rise to a DKS that features a mode-locked frequency comb spectrum with characteristic sech$^2$ spectral envelope\cite{kippenberg2018dissipative}. The microresonators employed here are dispersion-engineered ring resonators (microrings) with radii in the 30-60 $\mu$m range (Fig. 1a), fabricated on Z-cut TFLN (Z-TFLN) on insulator wafers (Methods). The microrings support fundamental transverse electric (TE) modes with intrinsic quality factors ($Q_i$) in the 1-2 million range, in the telecommunications C-band. The simulated and measured integrated dispersion ($D_{\text{int}}$) of a typical 50 $\mu$m-radius microring is shown in Fig. 1b. This microring is evanescently coupled to a bus waveguide using a pulley coupler, which leads to a modal dissipation profile $Q_c$ with strong frequency dependence, as shown in Fig. 1c (for detailed study of the $Q_c$ dependence on pulley coupler parameters, see Supplementary information).

A CW laser source, providing 100 mW of on-chip power, is used to pump one microring resonance resulting in single and multi-DKS steps in the generated comb power trace (Fig. 1d). Here, DKS states can be accessed by sweeping the pump laser in both directions, owing to the unique interplay between the photorefractive effect on one hand, and thermo-optic and Kerr effects on the other. Their combined effects also stabilize the laser-resonance detuning against pump frequency fluctuations in the DKS regime, enabling excellent free-running stability of DKSs on TFLN. An optical spectrum corresponding to the single-DKS (multi-DKS) step in Fig. 1d is shown in Fig. 1e (f), with comb spacing of about 410.15 GHz and comb span of about 110 (112.4) THz. Power dips in both spectra (around 203 THz) are due to the anti-phase-matched condition of the pulley coupler and correspond to the $Q_c$ peak in Fig. 1c. This coupling condition allows critical coupling of the pump mode and strong overcoupling of the Raman mode, thereby suppressing resonance-enhanced stimulated Raman scattering (SRS) in favor of four-wave-mixing (FWM) and DKS generation, which we discuss next.

\begin{figure*}[t!]
\centering
\includegraphics[width=\linewidth]{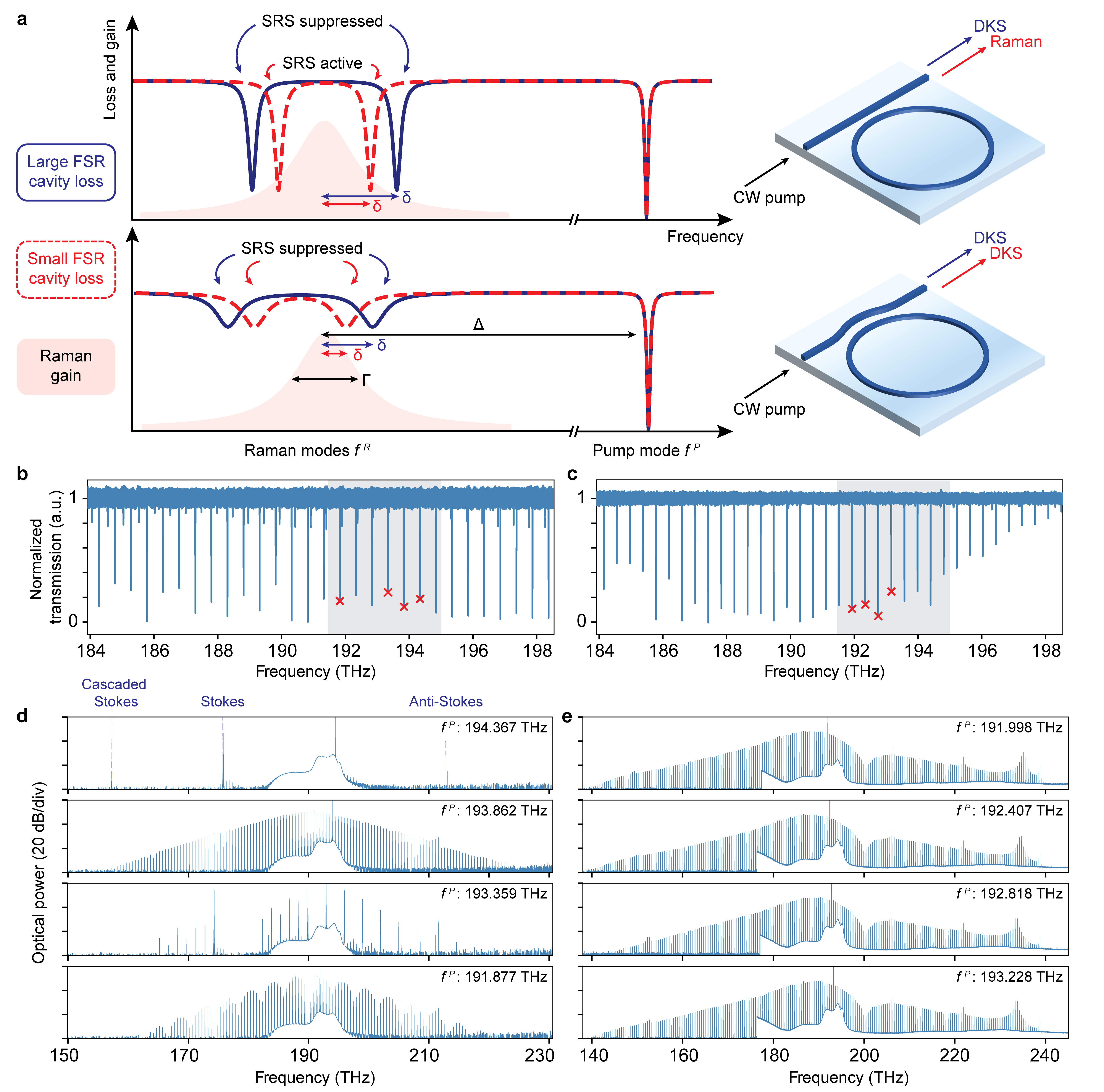}
\renewcommand\figurename{Fig.}
\caption{\label{fig:Fig2}\textbf{FSR control and dissipation engineering methods for SRS suppression.} \textbf{a}, Schematic of FSR control and dissipation engineering methods. In the FSR control method (top), empty microring modes are far detuned from the peak Raman gain (red shaded region), lowering the effective Raman gain coefficient. This method prefers large microring mode spacings (dark blue line) compared to the Raman gain bandwidth. Otherwise, small microring mode spacings (dashed red line) provide low loss conditions for the Raman gain which lowers the Raman lasing threshold. The dissipation engineering method (bottom), on the other hand, strongly overcouples modes near peak Raman gain and introduces significant loss. This leads to suppression of SRS irrespective of microring FSR. \textbf{b}, Transmission spectrum of an FSR-controlled, 40 $\mu$m-radius microring with FSR$\sim$491.7 GHz close to $\Gamma$ and satisfying $\Delta$/FSR$\sim$38.5. Shaded gray region indicates the bandwidth of the optical amplifier (191.5-194.7 THz range) used in the experiment. \textbf{c}, Transmission spectrum of a dissipation-engineered, 50 $\mu$m-radius microring with FSR$\sim$410.6 GHz. High frequency modes are undercoupled due to the $Q_c$ peak placed around 200 THz, and consequently, modes within the amplifier bandwidth are critically coupled while Raman modes are strongly overcoupled. \textbf{d}, \textbf{e}, Spectra generated by pumping modes marked by red crosses in \textbf{b}, \textbf{c}, respectively. In \textbf{d}, four modes generating distinct nonlinear states are selected (unmarked modes generate SRS), while in \textbf{e}, four consecutive modes supporting nearly identical DKS are shown (unmarked modes support DKS).}

\end{figure*}


\begin{figure}[t!]
\centering
\includegraphics[width=\linewidth]{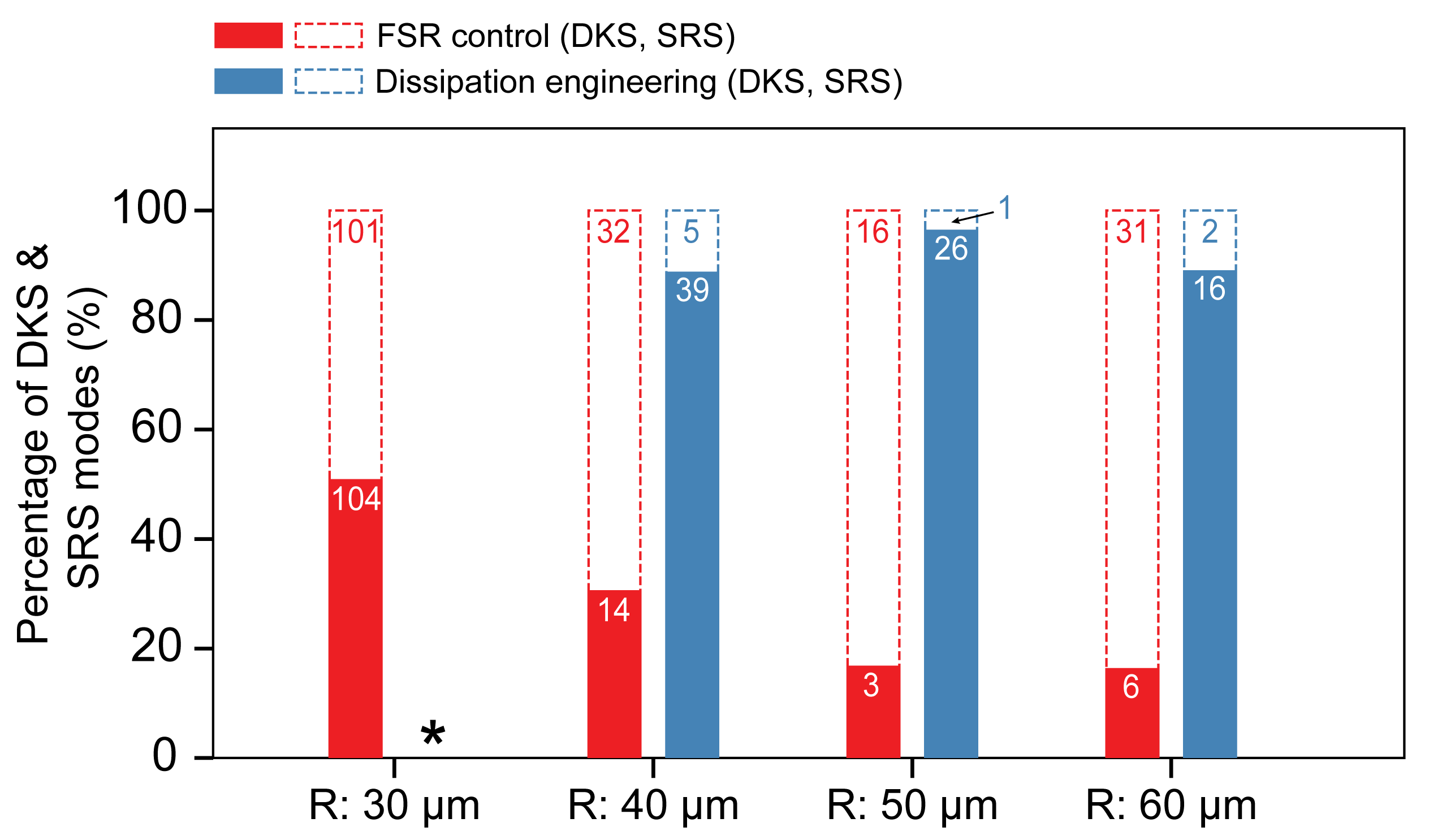}
\renewcommand\figurename{Fig.}
\caption{\label{fig:Fig3}\textbf{DKS yield.} Percentage of microring modes that yield DKS states (solid) or SRS (dashed), utilizing the FSR control method (red) and dissipation engineering method (blue). Microrings with radii 30, 40, 50, and 60 $\mu$m (FSRs of about 660, 492, 400, and 335 GHz, on average) were measured. Dissipation-engineered microrings with radius 30 $\mu$m were not attempted (star). If the comb power trace of a pump mode indicates clear evidence of DKS steps (as in Fig. 1b and Supplementary information), it contributes to a DKS count. Otherwise, it contributes to an SRS count. The number of modes resulting in DKS and SRS, for each microring radius and each SRS suppression method, is labeled. A total of 396 modes across 94 microrings have been tested, where all microrings selected satisfied design guidelines.}

\end{figure}

\vspace{1ex}\noindent\textbf{Resonantly enhanced SRS vs. Kerr effect}\\
Z-TFLN is known to feature strong SRS associated with the $\chi^{(3)}$-coupling between the fundamental TE mode and the crystalline E(LO$_8$) vibrational mode\cite{gong2019soliton}. As a CW pump enters a microring resonance mode from the red-detuned side, if the SRS threshold is lower than those of other nonlinear processes\cite{yu2020raman,zhao2023widely}, then it will prompt an immediate energy transfer from the pump mode to unoccupied resonance modes around the Raman gain band. Such Raman lasing behavior clamps the pump power at the SRS threshold provided Raman gain is not saturated, and it inhibits DKS generation despite appropriate pump-resonance detuning conditions. To enter DKS states, the Kerr-mediated FWM process which seeds sidebands must occur instead of SRS\cite{chembo2013spatiotemporal}. 

To qualitatively describe the competition between SRS and DKS formation, we consider the ratio between the resonance-enhanced power thresholds of these processes, given by
\begin{equation}
\zeta= \frac{P_{\text{th}}^{\text{Kerr}}}{P_{\text{th}}^{\text{SRS}}}\sim\frac{Q_L^R}{Q_L^P}\cdot\frac{g_{\text{eff}}^{\text{SRS}}(\delta)}{g^{\text{Kerr}}},
\end{equation}
where $P_{\text{th}}^{\text{Kerr}},P_{\text{th}}^{\text{SRS}}$ are the FWM and SRS power thresholds, $Q_L^P,Q_L^R$ are the loaded quality factors of the pump mode and Raman mode (Raman mode is defined as the microring resonance mode closest to the Raman gain peak located at $f^P-\Delta$, where $f^P$ is the pump frequency and $\Delta\sim18.94$ THz is the Raman shift\cite{basiev1999raman,ridah1997composition}), $g^{\text{Kerr}}$ is the FWM gain coefficient, $g_{\text{eff}}^{\text{SRS}}(\delta)$ is the Raman gain function, and $\delta=f^R-(f^P-\Delta)$ is the detuning between the Raman mode and the Raman gain peak $f^P-\Delta$. When $\zeta<1$, DKS formation is favored over SRS. However, since $g_{\text{eff}}^{\text{SRS}}(\delta=0)/g^{\text{Kerr}}\sim31\gg 1$, the $\zeta>1$ regime is typical and SRS is routinely observed. Importantly, $\zeta$ may be lowered by explicitly engineering $Q_L^P,Q_L^R,g_{\text{eff}}^{\text{SRS}}(\delta),$ and $g^{\text{Kerr}}$. Among these parameters, $g^{\text{Kerr}}=4\pi n_2 f^P/c$, where $c$ is the speed of light and $n_2$ is the nonlinear coefficient (proportional to $\chi^{(3)}$), is not easily tunable. On the other hand, microring resonance frequencies can circumvent the Raman gain band through careful mode placement and FSR control, such that $\delta$ is large for the Raman mode\cite{okawachi2017competition,gaeta2019photonic} and $g_{\text{eff}}^{\text{SRS}}(\delta)$ is reduced from its peak value. Alternatively, introducing a highly dissipative channel for the Raman mode while maintaining efficient coupling for the pump mode\cite{he2023high,gong2020photonic} reduces $Q_L^R/Q_L^P$ greatly. The former method we refer to as ``FSR control method" and the latter ``dissipation engineering method". Both methods effectively lower $\zeta$ and are conceptually described in Fig. 2, along with representative nonlinear states obtained by pumping four different modes in each case. Further, Fig. 3 statistically analyzes the efficacy of both methods for DKS generation and SRS suppression in Z-TFLN, by pumping 396 modes across 94 microrings with four distinct radii in the 30-60 $\mu$m range.

\begin{figure*}[t!]
\centering
\includegraphics[width=\linewidth]{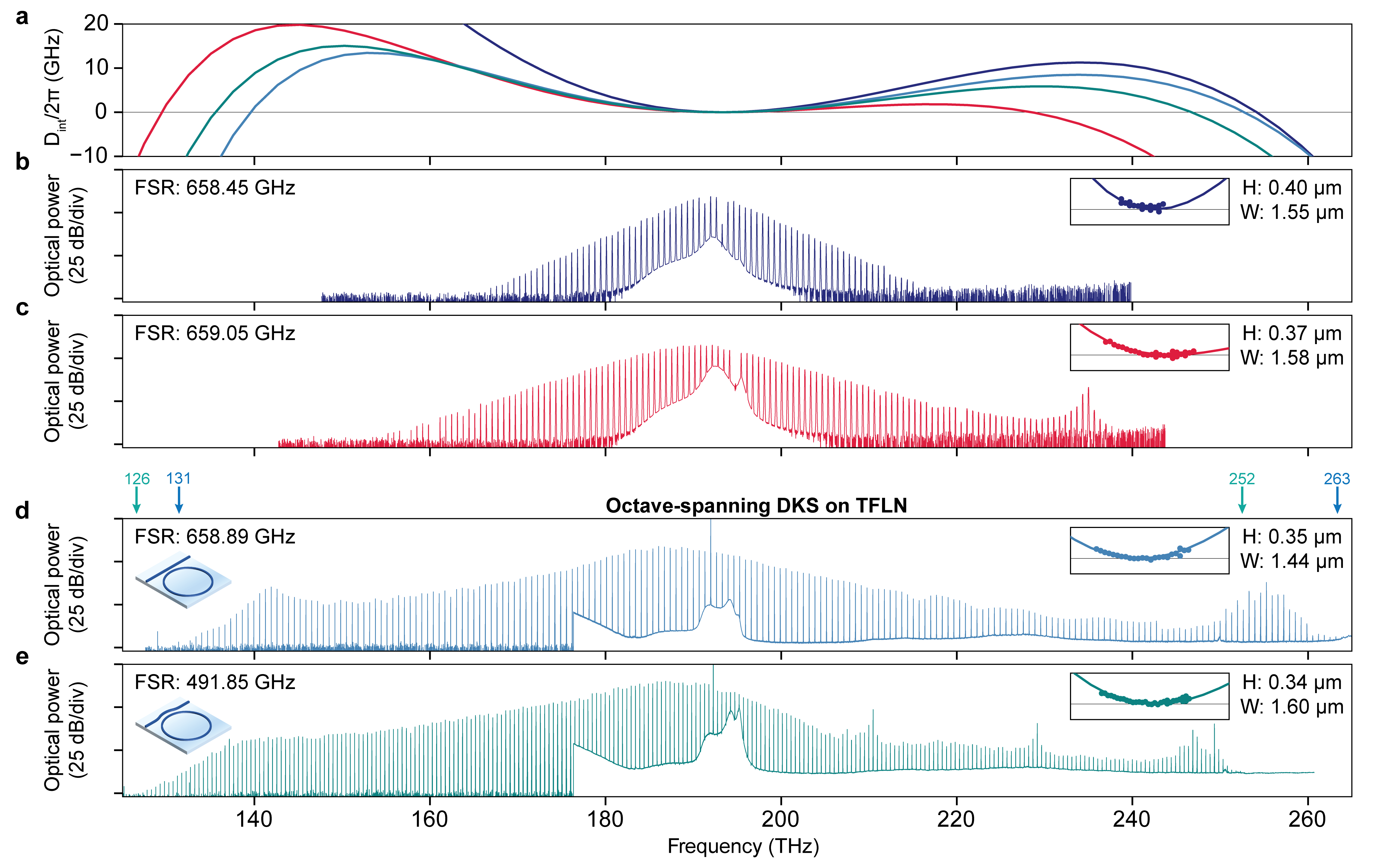}
\renewcommand\figurename{Fig.}
\caption{\label{fig:Fig4}\textbf{Dispersion engineering and octave-spanning DKS on TFLN.} \textbf{a}, Integrated dispersion ($D_{\text{int}}$) of four microring resonators. \textbf{b}, Single-DKS state from a 658.45 GHz FSR microring with large anomalous dispersion. No dispersive waves are produced at moderate pump powers of about 50 mW on-chip. \textbf{c}, Single-DKS state from a 659.05 GHz microring with reduced anomalous dispersion, compared to \textbf{b}. A single dispersive wave at high frequency is generated with a similar pump power. \textbf{d}, Octave-spanning single-DKS state (131 to 263 THz, labeled by blue arrows) from a 658.89 GHz FSR microring broadened by dual dispersive waves. This microring utilizes the FSR control method for DKS generation. \textbf{e}, Octave-spanning single-DKS state (126 to 252 THz, labeled by green arrows) from a 491.85 GHz FSR microring broadened by dual dispersive waves. A weak coupling resonance induced distortion in the spectral envelope is visible near 210 THz. This microring utilizes the dissipation engineering method for DKS generation.}
\end{figure*}

\vspace{1ex}\noindent\textbf{FSR control method}\\
In the FSR control method, we minimize spectral overlap between the Raman gain band and adjacent microring Raman modes, thereby tuning $\delta$ away from $0$ and reducing the effective Raman gain $g_{\text{eff}}^{\text{SRS}}(\delta)$. Optimized modal positions in the vicinity of the Raman gain are schematically illustrated in the top panel of Fig. 2a. The peak Raman gain is centered between two unoccupied modes, $\Delta$/FSR is a half-integer, and $\delta=$FSR$/2$. When $\delta$ is not large enough (red dashed line in Fig. 2a, top), Raman lasing conditions are fulfilled before FWM can occur. Since $\delta$ is proportional to FSR, the larger the FSR (blue solid line in Fig. 2a, top), the smaller the $g_{\text{eff}}^{\text{SRS}}(\delta)$ \cite{okawachi2011octave}. To date, this approach has been effective for Kerr comb generation in crystalline materials with small $\Gamma$, such as diamond $(\Gamma\sim60$ GHz) and silicon $(\Gamma\sim105$ GHz)\cite{okawachi2017competition,gaeta2019photonic}. In Z-TFLN, $\Gamma$ associated with the E(LO$_8$) vibrational mode is 558 GHz, thus a large FSR is required. This restricts the utility of the FSR control method since frequency combs with large FSR have limited usability. Further, large FSR requires rings with small radii ($<40$ $\mu$m), which are subject to increased bending losses resulting in larger thresholds   $P_{\text{th}}^{\text{Kerr}}\sim1/(Q_L^P)^2$, as well as from scattering-induced mode couplings that lead to $Q_L^P$ and $Q_L^R$ variations across different modes. This may be detrimental to the consistency of SRS suppression and renders this approach less deterministic, especially when FSR$\gg\Gamma$ cannot be achieved due to fabrication limitations.

To test this method, we fabricated a microring with 39.84 $\mu$m-radius and measured FSR of about 491.7 GHz, satisfying $\Delta$/FSR$\sim$38.5 and FSR$\sim\Gamma$. The transmission spectrum of this device is shown in Fig. 2b. Four resonance modes (red crosses) were separately pumped with about 100 mW on-chip power, and two out of four resonances yielded exclusively single or multi-DKS states, as shown in Fig. 2d. The other two resonances produced strong Raman lasing (single and cascaded Raman shifts of the pump light) and Raman-assisted modulation instability (Raman light coupled with chaotic sideband generation) which are not mode-locked frequency combs. Results collected across many microrings and their modes indicate that, while the FSR control method can produce DKS states, the outcome is hard to predict, and it is not robust in suppressing SRS on TFLN, down to the smallest microring dimensions that may still achieve low loss.

\vspace{1ex}\noindent\textbf{Dissipation engineering method}\\
In the dissipation engineering method, we utilized a pulley-shaped microring to bus waveguide coupler, which tailors the dissipation rates of microring modes near the Raman gain band relative to those within the C-band  (optical amplifier operating band). This  significantly increases the SRS threshold while maintaining a low FWM threshold for the pump. The amount of dissipation per mode, in addition to intrinsic material absorption and scattering losses (reflected by the intrinsic quality factors $Q_i^P$ and $Q_i^R$), is determined by the coupling to external channels which extracts energy from that mode (reflected by the coupling quality factors $Q_c^P$ and $Q_c^R$). The overall loaded quality factor is then $(Q_L)^{-1}=(Q_i)^{-1}+(Q_c)^{-1}$, which enters the expression for $\zeta$. Assuming $Q_i^P\sim Q_i^R$ (a good approximation since $f^P$ and $f^R$ are sufficiently close and far from the material band gap), we have $Q_L^R/Q_L^P\ll 1$ provided $Q_c^R/Q_c^P\ll 1$. If this ratio is small enough, it may offset $g_{\text{eff}}^{\text{SRS}}(\delta)/g^{\text{Kerr}}$ completely for arbitrary $\delta$, and enforce the DKS generation condition $\zeta<1$. This method thus requires engineering $Q_c^R$ relative to $Q_c^P$ such that $Q_c^R/Q_c^P\ll 1$. Importantly, standard point couplers provide $Q_c^R/Q_c^P\sim 1/5$ for a critically coupled pump, while pulley couplers enable $Q_c^R/Q_c^P\lesssim 1/40$ (Supplementary information). The latter is attributed to strongly frequency-dependent $Q_c$ near anti-phase matched coupling resonances (realized as in Fig. 1c)\cite{moille2019broadband}. When such coupling resonances are just higher frequency than the pump band (C-band), pump modes may be critically coupled while Raman modes are strongly overcoupled. Therefore, $\zeta<1$ is possible even considering the worst case $g_{\text{eff}}^{\text{SRS}}(\delta=0)/g^{\text{Kerr}}\sim 31$, thereby lifting all restrictions on the microring dimensions in principle. This is schematically illustrated in the bottom panel of Fig. 2a, where Raman mode loss cannot be balanced by Raman gain regardless of the FSR, and Raman lasing is always suppressed.

The transmission spectrum of a dissipation-engineered microring with 50 $\mu$m-radius is shown in Fig. 2c. It has a measured FSR about 410.6 GHz, resulting in  $\Delta$/FSR$\sim$49.2 and $\delta\sim0.2\cdot$FSR=82.2 GHz. We note that without dissipation engineering, such $\delta$ would likely result in SRS with high probability based on FSR control arguments. The undercoupled modes around 198 THz indicate presence of a $Q_c$ resonance just higher frequency than the C-band, which ensures critically coupled modes in the C-band and strongly overcoupled Raman modes, consistent with our dissipation engineering design strategy. Four consecutive resonances (red crosses) were separately pumped with about 100 mW on-chip power. All resonances yielded a single-DKS state, as shown in Fig 2e. In fact, these resonances also support many multi-DKS steps, indicating suppressed SRS over a wide range of pump-resonance detuning. Applying this method, a much more complex DKS phase space can be revealed, and multi-DKS states with higher pump-to-comb conversion efficiencies are readily accessible on TFLN.

\vspace{1ex}
\noindent\textbf{Comparison of methods and DKS yield}\\ 
To carefully verify our proposed methods and evaluate DKS yield statistics, we tested 94 dispersion-engineered microrings (applying either method for SRS suppression) and a total of 396 modes across these rings, as shown in Fig. 3. We confirm that engineering the microring dissipation spectrum can significantly increase DKS yield up to 96 \% (50 $\mu$m-radii) and greater than 88 \% for all other microring dimensions tested. The deviation from 100 \% yield can be attributed to (i) pumping near mode crossings or (ii) $Q_c$ sensitivity to frequency, where in some microrings a few modes in the amplifier bandwidth were undercoupled (and did not yield DKS) while the rest of the modes were critically coupled (and yielded DKS). In the FSR control method, we note that occasional DKS generation is possible even when the FSR is a fraction of $\Gamma$, or when the half-integer condition is imperfect resulting in $\delta<$FSR/2. This may be explained by dissipation rates of the fundamental TE modes further altered by scattering-induced mode coupling and possibly crossings between mode families, which in turn decreases $Q_L^R/Q_L^P$ and lowers $\zeta$, though in an inconsistent way.


\vspace{1ex}\noindent\textbf{Dispersion engineering and octave-spanning DKS}\\
Finally, in conjunction with the SRS suppression methods developed, we engineer the microring waveguide cross section by controlling the waveguide height, width, sidewall angle, bending radius, and cladding properties to achieve $D_{\text{int}}$ suitable for octave-spanning DKS (Fig. 4a). DKS states and their spectra are determined by $D_{\text{int}}(\mu)$, where $\mu$ denotes the comb line number relative to the pump mode ($\mu=0$), expressed as $D_{\text{int}}(\mu)=\sum_{n=2}^\infty\frac{D_n}{n!}\mu^n$. When the pump experiences small anomalous dispersion ($D_2\gtrsim0$), higher order dispersions ($D_n$ for $n\geq 3$) are non-negligible and responsible for broadening single- and multi-DKS states into frequencies of normal dispersion\cite{brasch2016photonic}, significantly extending the comb bandwidth through dispersive wave emissions at $\mu_{\text{DW}}$ where $D_{\text{int}}(\mu_{\text{DW}})\sim0$. Using the FSR control method, we fabricated three microrings with 30.3 $\mu$m-radius resulting in measured FSRs in the 658-659 GHz range (Fig. 4b-d). While $\Delta$/FSR$\sim$28.7 is not perfect half integer, we found the reduction in $g_{\text{eff}}^{\text{SRS}}(\delta)$ to be sufficient in practice, owing to the large FSR. For these devices, we showcase a DKS with sech$^2$ spectrum, a DKS broadened by one dispersive wave, and an octave-spanning DKS (131-263 THz, pumped with about 363 mW on-chip power) broadened by dual dispersive waves. Using the dissipation engineering method, we fabricated a pulley-coupled microring with measured FSR of 491.85 GHz, which also supports an octave-spanning DKS (126-252 THz, pumped with about 375 mW on-chip power).

\section{Discussion}
In summary, we demonstrated dispersion-engineered DKS states spanning up to an octave on Z-TFLN, using two distinct methods of microring FSR control and dissipation engineering. Our work is complemented by a statistically-validated understanding of DKS generation from TFLN microrings, as opposed to production of Raman scattered light. With symmetrized dispersion profiles about the pump frequency, DKS spectral bandwidths may be further improved by extending dispersive waves to be an octave apart, simplifying carrier envelope offset frequency detection and stabilization by offering powerful comb lines with high signal-to-noise ratio. Their powers may be further enhanced with the self-balancing effect\cite{moille2023kerr} and may generate even stronger beating signals for $f$-2$f$ self-referencing. Such an improvement in comb span would require microring waveguides with larger anomalous dispersion. This is necessarily accompanied by a higher DKS threshold, which must be offset by improving $Q_L^P$ and thus $Q_i^P$. The $Q_i^P$ typical to microring dimensions in this work is between 1 and 2 million for modes in the C-band, mainly limited by initial 600 nm Z-TFLN requiring shallow etched waveguides for optimal $D_{\text{int}}$. Simulations suggest that film thicknesses between 480 to 500 nm allow fully etched waveguides to yield octave-spanning $D_{\text{int}}$ (Supplementary information), while a recent report has shown the tight confinement provided by ridge waveguides on Z-TFLN may increase $Q_i^P$ up to 4.9 million for microring dimensions considered in our work\cite{gao2023compact}. Further, the on- and off-chip facet losses may be reduced from 6 dB per facet in our proof of concept demonstrations (single mode waveguides exposed through manual cleaving) to 1.7 dB per facet using inverse taper couplers\cite{he2019low}, greatly improving DKS extraction from the chip, as well as reducing external pump power requirements to be compatible with butt-coupled distributed feedback laser sources for further integration. Further, we utilized a fabrication workflow based on iterative dry/wet etching of TFLN, providing control over the microring waveguide sidewall angle. This additional degree-of-freedom for dispersion engineering expands the dispersion landscape and may enable a variety of nonlinear optics realized on TFLN, such as soliton quiet point\cite{stone2020harnessing}, parametrically driven solitons\cite{moille2024parametrically}, Raman solitons\cite{yang2017stokes,li2024ultrashort}, and so on.
In addition to FSR control and dissipation engineering methods, mode crossings and other spectral defects that may locally lower $Q_L^R$ were not explored here, but may have contributed to DKS generation. Such features can be explicitly engineered through matching effective indices between the fundamental TE pump and other mode families at the Raman-shifted frequency, though any detrimental effects on $D_{\text{int}}(\mu)$ must be evaluated. Defect mode engineering utilizing photonic crystal microrings may achieve similar ends, where sidewall corrugations strongly couple oppositely propagating modes at specific frequencies\cite{lu2020universal,yu2021spontaneous,zhang2024spectral} and may induce high loss at a mode near $f^P-\Delta$.

The DKSs demonstrated and methods introduced enable direct integration of octave-spanning optical frequency combs with existing advancements on the TFLN platform. Combined with high-speed electro-optic modulation and efficient second harmonic generation on-chip, fully stabilized integrated frequency standards are realizable through the detection of near-THz comb spacings using electro-optic down-conversion and detection of carrier envelope offset frequency using octave-separated comb lines coupled with $f$-2$f$ interferometry. Towards this goal, reproducible fabrication of the octave-spanning DKS source is critical to benchmarking such a large-scale system combining various photonic integrated components, also a core proponent to our work which may accelerate future system development. Such systems would provide a complete microwave to optical link with significant potential for application. For example, making use of integrated laser technology\cite{de2021iii,han2021electrically,shams2022electrically,snigirev2023ultrafast}, small form factor optical frequency synthesizers and frequency-precise spectroscopic light probes are possible. Considering the mutual synergies of second- and third-order nonlinearities on TFLN for comb generation and stabilization, we envision chip-scale TFLN frequency comb systems to meet technology’s growing needs for compact generators of an equidistant grid of ultra-stable and mutually coherent optical frequencies.

\bibliography{Reference}

\clearpage
\section{Materials and methods}
\noindent\textbf{Device fabrication}   \\
Dissipative Kerr soliton (DKS) microrings are fabricated on commercial Z-cut thin-film lithium niobate (Z-TFLN) on insulator wafers (NanoLN). The wafer stack consists of 600 nm Z-TFLN and 2 $\mu$m thermal oxide atop a 0.525 mm silicon handle. Bus and microring waveguides are patterned on hydrogen silsesquioxane (HSQ) resist using electron-beam lithography (EBL). An optimized etching process alternating Ar$^+$-based reactive ion etching and wet etching in SC-1 solution is utilized to steepen waveguide sidewalls ($72\degree$ used in our devices), which symmetrizes the anomalous dispersion profile about the pump frequency. Periodic wet etching removes redeposition buildup acting as an effective etch mask, which is responsible for typical shallow sidewalls below $60\degree$ (without angled etching). The HSQ resist is stripped with dilute hydrogen-fluoride (HF) solution and the devices are annealed in a high-temperature, oxygen-rich environment. Finally, bus waveguides are exposed through manual cleaving, resulting in 6 dB loss per facet, on average. We note that our fabrication process reduces device dimensions compared to their initial design, due to extensive wet etching. For integrated dispersion ($D_{\text{int}}$) comparisons, the measured $D_{\text{int}}$ for an as-designed microring waveguide top width is compared with a simulation assuming a 0.1 $\mu$m-reduced top width. For $Q_c$ comparisons, we rely on experimentally fine-tuning the bus waveguide width and coupling gap around a simulated optimal point.

\vspace{1ex}\noindent\textbf{Dispersion simulation and measurement} \\
Integrated dispersion $D_{\text{int}}(\mu)=\sum_{n=2}^\infty\frac{D_n}{n!}\mu^n$ of a microring waveguide is simulated by computing the eigenmodes of its cross section, using a commercial eigenmode solver (Lumerical MODE). Broadband effective index information is obtained which is used to calculate $D_{\text{int}}(\mu)$. The assumed microring waveguide sidewall angle is $72\degree$, using the etching process described above. The $D_{\text{int}}(\mu)$ of a microring is measured experimentally by scanning a tunable, C/L-band external cavity diode laser (ECDL) and obtaining the microring resonance mode positions. Such transmission spectra are normalized, and their frequency axes are calibrated using a fiber-based Mach-Zehnder interferometer (MZI) reference (191.3 MHz fringe spacing near 1.55 $\mu$m). This fringe spacing is calibrated against a frequency difference generated by a near 1.55 $\mu$m carrier frequency and its electro-optic sidebands. The resonance positions in the C/L-band are located, and the fitted FSR ($D_1$) at the pump frequency is removed. Such a measurement gives access to the experimental $D_{\text{int}}(\mu)$ expanded about $\mu=0$ and enables direct comparison against simulation.

\vspace{1ex}\noindent\textbf{DKS generation experiment} \\
DKS states are generated by a single-tone, continuous-wave pump (C-band ECDL) amplified by an erbium-doped fiber amplifier (EDFA). The amplified pump passes a polarization controller and is coupled onto the Z-TFLN chip by a lensed fiber. A 10 Hz electrical ramp is fed into the ECDL’s piezo-actuated frequency control to locate microring resonances and map comb power vs. pump-resonance detuning (Fig. 1b and Supplementary information). Once DKS steps are identified, the electrical ramp signal is turned off and the pump frequency is manually tuned into the DKS existence range starting from blue side of resonance. Note that a self-starting DKS may occasionally be triggered with an initially red-detuned pump frequency. Once the DKS state is entered, no additional locking mechanism is required. The generated comb spectra are collected by a lensed fiber and detected using two optical spectrum analyzers (Yokogawa AQ6370D, AQ6375). A tunable fiber Bragg grating (FBG) filter is used to remove the pump frequency when monitoring comb power or taking select spectra.

\vspace{1ex}\noindent\textbf{SRS suppression calculation} \\
SRS is a parasitic nonlinear process in TFLN microrings inhibiting DKS generation. We proposed two strategies to bias microrings in favor of DKS generation, based on Raman mode placement in conjunction with large FSR and dissipation engineering of Raman modes. Such strategies are conceived by comparing the threshold powers of SRS against four-wave-mixing (FWM):
\begin{equation}
P_{\text{th}}^{\text{SRS}}=\frac{\pi^2 n_0^2 f^P f^R}{c^2 g_{\text{eff}}^{\text{SRS}}(\delta)}\cdot \frac{V_{\text{eff}}}{Q_L^R Q_L^P}
\end{equation}
and
\begin{equation}
P_{\text{th}}^{\text{Kerr}}=\frac{\pi^2 n_0^2 (f^P)^2}{c^2 g^{\text{Kerr}}}\cdot \frac{V_{\text{eff}}}{(Q_L^P)^2},    
\end{equation}
where $n_0$ is the pump mode effective index, $V_{\text{eff}}=2\pi RA_{\text{eff}}$ is the pump mode effective mode volume, $A_{\text{eff}}$ is the pump mode effective mode area, $R$ is the microring radius, $f^R,f^P$ are the Raman and pump mode frequencies, $g^{\text{Kerr}}=4\pi n_2 f^P/c\sim0.146$ cm/GW is the FWM gain coefficient, $n_2=1.8\cdot10^{-19}$ m$^2$/W is the nonlinear index, $g_{\text{eff}}^{\text{SRS}}(\delta)$ is the effective SRS gain function for the E(LO$_8$) vibrational mode (peak gain $g_{\text{eff}}^{\text{SRS}}(\delta=0)\sim4.51$ cm/GW, scaled to 1.55 $\mu$m using measured values at 1 $\mu$m\cite{johnston1968stimulated}), and $\delta$ is the detuning between peak SRS gain and the nearest microring resonance mode. The $g_{\text{eff}}^{\text{SRS}}(\delta)$ has a center $\Delta\sim$18.94 THz away from the pump and a bandwidth $\Gamma\sim$558 GHz\cite{basiev1999raman,ridah1997composition}. A ratio of
\begin{equation}
\zeta=\frac{P_{\text{th}}^{\text{Kerr}}}{P_{\text{th}}^{\text{SRS}}}\sim\frac{Q_L^R}{Q_L^P}\cdot\frac{g_{\text{eff}}^{\text{SRS}}(\delta)}{g^{\text{Kerr}}}<1    
\end{equation}
supports DKS generation over SRS. Without special considerations, $\zeta>1$ for the fundamental transverse-electric (TE) mode at C-band frequencies in Z-TFLN. The FSR control and dissipation engineering methods tune the terms in $\zeta$ so that $\zeta$ is maximally decreased.

\vspace{1ex}\noindent\textbf{Coupler simulation for dissipation engineering} \\
The microresonator coupling is defined by the coupling rate $\kappa_c$ between the microresonator and the bus waveguide. The structure which facilitates evanescent coupling is called the coupler. The $\kappa_c$ of the coupler is represented by $\kappa_c=|t|^2\cdot c/(n_g R)$, where $t$ is the cross-coupling transmission of the coupler, $c$ is the speed of light, $n_g$ is the group index, and $R$ is the microresonator radius. Coupler design is important for DKS generation on TFLN as parasitic SRS, also based on the third-order nonlinear optical response, has a stronger gain coefficient (in most cases) than the Kerr nonlinear gain. The dissipation engineering method employs pulley couplers to increase the SRS threshold and suppresses Raman lasing in favor of DKS generation at all optical pump powers. These couplers are designed such that the pump frequency is critically coupled (in terms of coupling rates, $\kappa_i (f^P)\sim\kappa_c(f^P)$) while the Raman mode $f^R$ is strongly overcoupled compared to the pump $\kappa_c(f^P)\ll \kappa_c(f^R)$ and $ \kappa_i(f^P)\sim\kappa_i(f^R)\ll\kappa_c(f^R)$. This set of conditions provides efficient resonance enhancement of the pump field while increasing the lasing threshold for microresonator modes near $f^R$, favoring DKS generation. The design of these couplers thus focuses on the $\kappa_c(f^P)$ and $\kappa_c(f^R)$ contrast arising from strongly frequency-dependent $t$, where $t$s are directly accessed using 3-D electromagnetics simulations of the coupling structure using a commercial finite-difference time-domain solver (Flexcompute Tidy3D). Coupling quality factors $Q_c(f)=2\pi f/\kappa_c(f)$ are inversely proportional to the coupling rates.\\

\noindent\textbf{Data availability}  The data that support the plots within this paper and other findings of this study are available from the corresponding authors upon reasonable request.

\noindent\textbf{Code availability}  The code used to produce the plots within this paper is available from the corresponding authors upon reasonable request.

\noindent\textbf{Competing interests}  M.L. is involved in developing lithium niobate technologies at HyperLight Corporation.

\noindent\textbf{Note} During the manuscript review stage, we became aware of another work on an octave-spanning DKS on TFLN\cite{wang2024octave}.

\section*{Acknowledgments}
This work is supported by the Defense Advanced Research Projects Agency (HR001120C0137, D23AP00251-00), Office of Naval Research (N00014-22-C-1041), National Science Foundation (OMA-2137723, OMA-2138068), U.S. Navy (N68335-22-C-0413), and National Research Foundation of Korea. The authors thank Pradyoth Shandilya, Rebecca Cheng, and Neil Sinclair for discussions. Y.S. acknowledges support from the AWS Generation Q Fund at the Harvard Quantum Initiative. All devices in this work were fabricated at the Harvard Center for Nanoscale Systems.




\end{document}